\documentclass[twocolumn,article,9pt,superscriptaddress,showkeys,showpacs] {revtex4}
\usepackage{amsfonts}
\usepackage{graphicx, amsmath}
\begin{document}
	
	\title[Short Title]{Optical nonreciprocal response and conversion in a Tavis-Cummings coupling optomechanical system}
	\author{Yang Jiao}
	\affiliation{Department of Physics, College of Science, Yanbian University, Yanji, Jilin 133002, China}
	\author{Cheng-Hua Bai}
	\affiliation{School of Physics, Harbin Institute of Technology, Harbin, Heilongjiang 150001, China}
	\author{Dong-Yang Wang}
	\affiliation{School of Physics, Harbin Institute of Technology, Harbin, Heilongjiang 150001, China}
	\author{Shou Zhang\footnote{E-mail: szhang@ybu.edu.cn}}
	\affiliation{Department of Physics, College of Science, Yanbian University, Yanji, Jilin 133002, China}
	\author{Hong-Fu Wang\footnote{E-mail: hfwang@ybu.edu.cn}}
	\affiliation{Department of Physics, College of Science, Yanbian University, Yanji, Jilin 133002, China}
	\begin{abstract}
		We propose a scheme to realize optical nonreciprocal response and conversion in a Tavis-Cummings coupling optomechanical system, where a single cavity mode interacts with the vibrational mode of a flexible membrane with an embedded ensemble of two-level quantum emitters. Due to the introduction of the Tavis-Cummings interaction, we find that the phases between the mechanical mode and the optical mode, as well as between the mechanical mode and the dopant mode, are correlated with each other, and further give the analytical relationship between them.  By optimizing the system parameters, especially the relative phase between two paths, the optimal nonreciprocal response can be achieved. Under the frequency domain, we derive the transmission matrix of the system analytically based on the input-output relation and study the influence of the system parameters on the nonreciprocal response of the quantum input signal. Moreover, compared with the conventional optomechanical systems, the Tavis-Cummings coupling optomechanical system exhibits richer nonreciprocal conversion phenomena among the optical mode, mechanical mode, and dopant mode, which provide a new applicable way of achieving the phonon-photon transducer and the optomechanical circulator in future practice.
	\end{abstract}
	
	\maketitle \section{Introduction}\label{sec0}
	In recent years, cavity optomechanics~\cite{Aspelmeyer2014,Physics.2.40,Science.321.1172} (COM) has attracted a lot of attention due to its theoretical and experimental rapid development, which has produced many interesting phenomena, e.g., the detection of gravitational waves~\cite{2018AAPPS}, mechanical squeezing~\cite{Bai2019a,2008APL92133102,Han2019,2008NP4785,Bai2019,2016SR638559}, quantum entanglement~\cite{PRL10,PRA16}, mechanical cooling~\cite{Liu2018,PRL21,Wang2018a}, nonclassical correlations between single photon and phonon~\cite{N22}, photon blockade~\cite{Wang2019}, and coherent wavelength conversion~\cite{S23,N24}. The common optomechanical systems study the two-body interaction between the cavity field and mechanical resonator. A novel interaction has been studied lately, which is the three-body interaction among the cavity field, mechanical resonator, and atom. Such hybrid systems are called Tavis-Cummings coupling optomechanical system, which has attracted wide attention. In the hybrid system, the cavity field interacts with mechanical resonator and few-level system simultaneously, which can lead to quantum interference effects and amazing optomechanical phenomena, for instance, EIT and EIA~\cite{2019OE277344,2020OE28580,2015OE2311508}, sensing~\cite{2019PR71440,2019QE1e10}, quantum repeater~\cite{2012PRA85062311}, interaction with an atomic ensemble may be used to produce entanglement, optomechanical cooling, backaction evading measurements of mechanical motion~\cite{2007IJTP462550,Dantan2014,Q73,2007PRA75022312}.
	
	On the other hand, COM-based optical nonreciprocal phenomena have also attracted significant interest in the decades, which means that the transmission of signals in two opposite directions exhibits different characteristics. Optical nonreciprocity has been realized in various optomechanical structures~\cite{SR43,OE44,F64}, such as magneto-optical crystals~\cite{APL45,PRL46,Li2016}, optical nonlinear systems~\cite{OL50,S51,QEe112019}, spatial-symmetry-breaking structures~\cite{APL52,OE53,OE54}, parity-time-symmetric structures~\cite{PRA55,Guo2014,2014NP105394}. Nonreciprocal devices based on nonreciprocity, e.g., isolator, directional amplifier, circulator, etc, play significantly important roles in quantum information processing and communication. Specifically, optical isolator is a device that blocks light in one direction but allows light to pass in the opposite direction, which has already been studied in both theory~\cite{2017AO562991} and experiment~\cite{OE32,NP27}. The fabrication of low-loss and high Q optical microcavities has achieved great progress using chemo-mechanical polish lithography~\cite{2019QE1e9}. The directional amplification has also been studied in a double-cavity optomechanical system with mechanical gain~\cite{PRA33}, a superconducting microwave circuit with parametric pumps~\cite{PRX34,PRA35}, a triple-cavity optomechanical system with optical gain~\cite{PRA36}, and a three-mode optomechanical system with an additional mechanical drive~\cite{OE37}. It has also been realized in an optomechanical circuit via synthetic magnetism and reservoir engineering in experiment~\cite{NP38}. Circulator has been investigated in three-mode optomechanical systems~\cite{PRA39} and this technique can be applied to a circuit-QED architecture~\cite{PRA40,Qi2018} and phonon devices~\cite{NJP41}.
	
	As we all know, the nonreciprocal effect relies strongly on the relative phase between two different paths, which can be usually reduced to the phase at a single path in most theoretical researches. When the phase satisfies the condition of destructive interference, the nonreciprocal phenomenon occurs, which is similar to the optomechanically induced transparency~\cite{Liu2015,AP2018,Zheng2019}. The optical nonreciprocal effect can be achieved by the momentum difference between the forward and backward moving light beams in an optomechanical system~\cite{PRL58}. Li $et~al$. proposed a multimode system consisting of two vibrating membranes suspended inside two cavities, which can produce the tunable optical nonreciprocity and be used as a phonon-photon router~\cite{PRA59}. Hafezi $et~al$. pointed out that the nonreciprocal transmission can be achieved via using an unidirectional optical pump in a microring resonator~\cite{OE60}. We also note that the optical nonreciprocity is studied in coupled spinning optomechanical resonators~\cite{OE63}. Furthermore, nonreciprocal response has been realized in experiments until now, for instance, silicon chip~\cite{PRL72},  strong dispersive coupling of a high-finesse cavity to a micromechanical membrane~\cite{N65}, a dispersive optomechanical system consisting of a membrane inside a cavity~\cite{NJP66}, and nonlinear coupling in a low-loss system~\cite{NP67}. Meanwhile, some other studies related to nonreciprocity have been reported such as fast-slow light effects~\cite{OE61} and nonreciprocity photon blockade~\cite{PRL71}. On the other hand, the hybrid phonon-photon systems have been studied extensively for the interesting nature due to their potential applications in information processing network~\cite{2014PRA90023817,Xing2018,2019PRA99013804}. It has been shown that the conversion between phonons and photons can be realized and the optomechanical circulator can be engineered in the optomechanical systems~\cite{NJP1}.
	
	In this paper, we study the nonreciprocal response and conversion effects in a Tavis-Cummings coupling optomechanical system, which is composed of a trapped flexible membrane embedded an ensemble of two-level quantum emitters in an optical cavity. Resorting to the general linearization technique and the Fourier transform, we calculate the transmission matrix analytically and study the nonreciprocal response of the fluctuation signal in the frequency domain. In the present proposal, due to the introduction of the Tavis-Cummings coupling, we find that the phases of the two different paths are correlated each other and further we derive their relation analytically, which is greatly different from the previous studies. By selecting the system parameters appropriately, the prefect nonreciprocal response can be achieved. We also show that the nonreciprocity induces the feasibility to perform signal conversion among the optical mode, mechanical mode, and dopant mode, which implies that the system can be applied as a phonon-photon transducer and an optomechanical circulator. These interesting phenomena indicate that the Tavis-Cummings coupling model has potential applications in experiments, such as integrating the nonreciprocal devices.    
	
	The organization of the paper is as follows. In Sec.~II, we present the Hamiltonian and dynamical equations of the Tavis-Cummings coupling optomechanical system. In Sec.~III, we discuss the effect of the system parameters on nonreciprocal response in detail and show that the system can be applied as a phonon-photon transducer and an optomechanical circulator. Finally, we draw the conclusion in Sec.~IV.  
	
	\section{Model and Hamiltonian}\label{sec1}
	\begin{figure}
		\includegraphics[width=3.3in]{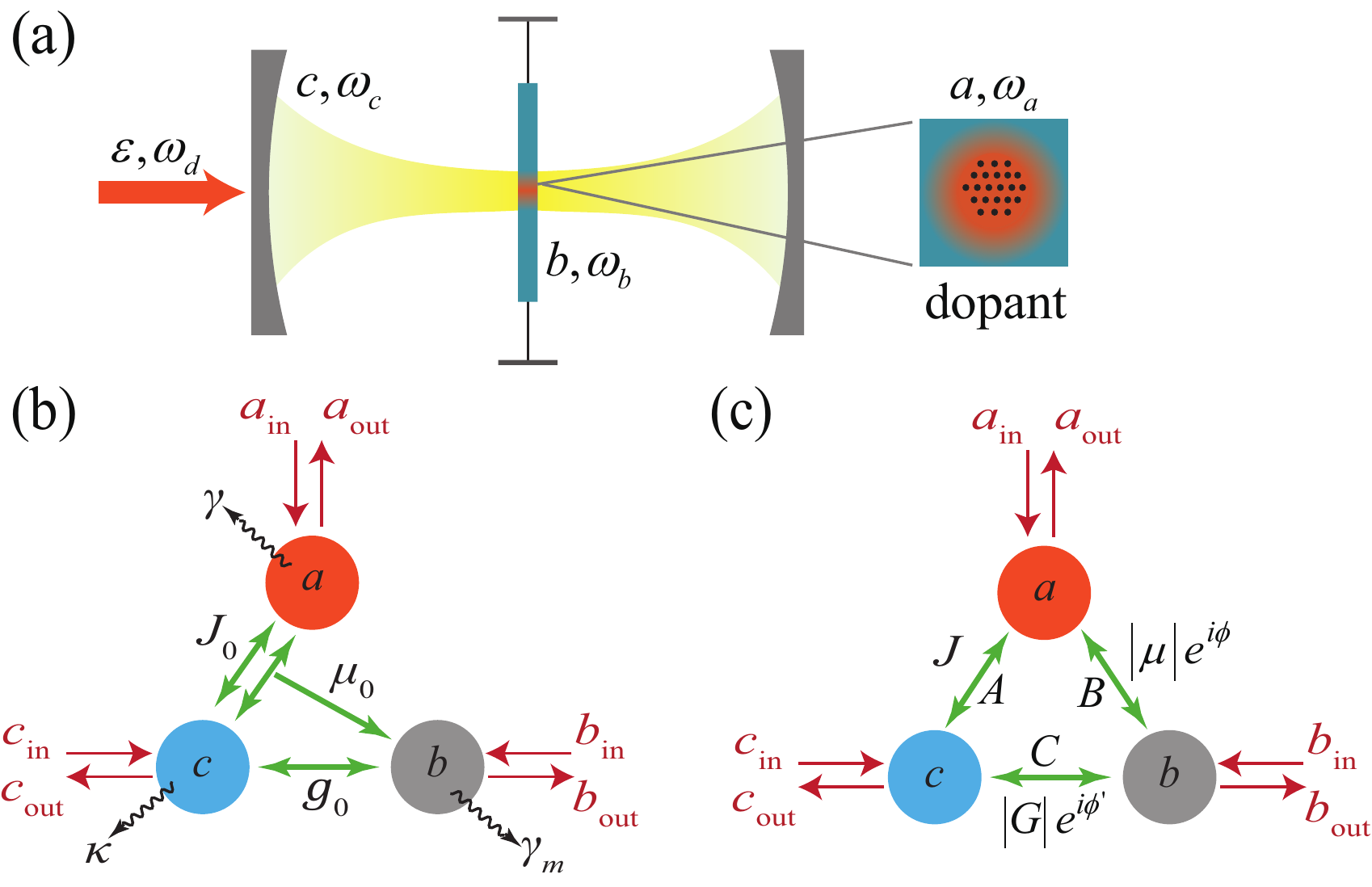}
		\centering
		\caption{(Color online) (a) Schematic of the Tavis-cummings coupling optomechanical setup, in which a cavity mode interacts with a single vibrational mode of a flexible membrane with an embedded ensemble of two-level quantum emitters that collectively behave as a single bosonic mode. (b) Depiction of the interaction of the Tavis-Cummings coupling optomechanical system given by Hamiltonian in Eq.~(\ref{e02}). (c) The effective interactions in the linearized regime of the system~\cite{Dantan2014,Q73}.}
		\label{fig1}
	\end{figure}
	
	As depicted in Fig.~\ref{fig1}(a), we consider a Tavis-cummings coupling optomechanical system, where an optical cavity mode interacts with a single vibrational mode of a flexible membrane with an embedded ensemble of two-level quantum emitters. Generally speaking, the operator of two-level quantum emitters is represented by the Pauli operator, here we represent the Pauli operator as a bosonic operator by a bosonic transformation. Under the conditions of sufficiently large two-level quantum emitters number $N$ and the limit of weak excitation, the lowering operators (raising operators) $\sigma_{-}~(\sigma_{+})$ of the ensemble can be transformed to a collective bosonic operator $a~(a^{\dagger})$ in the Holstein-Primakoff representation,
	\begin{eqnarray}\label{e01}
	\sigma_{-}&=&a\sqrt{N-a^{\dagger}a}\approx\sqrt{N}a,\cr\cr
	\sigma_{+}&=&a^{\dagger}\sqrt{N-a^{\dagger}a}\approx\sqrt{N}a^{\dagger},\cr\cr
	\sigma_{Z}&=&a^{\dagger}a-\frac{N}{2},
	\end{eqnarray}
	where operators $a$ and $a^{\dag}$ obey the bosonic commutation relation [$a,a^{\dag}$] = 1~\cite{Dantan2014,Wang2016,2015PRA92033841}. The total Hamiltonian of the system is written as ({$\hbar$} = 1)
	\begin{eqnarray}\label{e02}
	H&=&H_{0}+H_{\rm{int}}+H_{\rm{dr}},
	\end{eqnarray}
	with 
	\begin{eqnarray}\label{e03}
	H_{0}&=&\omega_{c}c^{\dag}c+\omega_{a}a^{\dag}a+\omega_{m}b^{\dag}b,\cr\cr
	H_{\rm{int}}&=&g_{0}c^{\dag}c(b^{\dag}+b)+[J_{0}+\mu_{0}(b^{\dag}+b)](a^{\dag}c+c^{\dag}a),\cr\cr
	H_{\rm{dr}}&=&i\varepsilon(c^{\dag}e^{-i\omega_{d}t}-\mathrm{H.c.}),
	\end{eqnarray}
	where $H_{0}$ describes the free Hamiltonian of the system, $H_{\rm{int}}$ represents the interactions of the cavity field with the mechanical membrane via radiation pressure and with the dopant (the embedded ensemble of two-level quantum emitters) via a mechanically modulated Tavis-Cummings coupling, and $H_{\rm{dr}}$ denotes the interaction between the driving field and the cavity field. Here, $c$~($c^{\dag}$) is the annihilation (creation) operator of the cavity mode with frequency $\omega_{c}$, $a$~($a^{\dag}$) is  the annihilation (creation) operator of the dopant mode with frequency $\omega_{a}$, and $b$~($b^{\dag}$) is the annihilation (creation) operator of the mechanical mode with frequency $\omega_{m}$. $g_{0}$ is the optomechanical coupling strength between the cavity mode and the mechanical mode, $J_{0}$ is the coupling strength between the cavity mode and the dopant mode, and $\mu_{0}$ is the Tavis-Cummings coupling strength among the cavity mode, mechanical mode, and dopant mode. $\varepsilon$ and $\omega_{d}$ are the amplitude and frequency of the driving field, respectively. In Fig.~\ref{fig1}(b), we give the explicit schematic for describing the interaction of the proposed Tavis-Cummings coupling optomechanical system.
	
	In the rotation frame with $H_{r}=\omega_{d}(c^{\dag}c+a^{\dag}a)$, according to the Heisenberg equations of motion, we can get the quantum Langevin equations (QLEs)
	\begin{eqnarray}\label{e04}
	\dot{c}&=&-(i\Delta_{c}+\kappa)c-ig_{0}c(b^{\dag}+b)-i[J_{0}+\mu_{0}(b^{\dag}+b)]a\cr\cr
	&&+\varepsilon+\sqrt{2\kappa}c_{\rm{in}},\cr\cr
	\dot{a}&=&-(i\Delta_{a}+\gamma)a-i[J_{0}+\mu_{0}(b^{\dag}+b)]c+\sqrt{2\gamma}a_{\rm{in}},\cr\cr
	\dot{b}&=&-(i\omega_{m}+\gamma_{m})b-ig_{0}c^{\dag}c-i\mu_{0}(a^{\dag}c+c^{\dag}a)\cr\cr
	&&+\sqrt{2\gamma_{m}}b_{\rm{in}},
	\end{eqnarray}
	where $\Delta_{i}=\omega_{i}-\omega_d$ is the detuning of the respective mode $(i=c,a)$ from the driving frequency. $\kappa$ ($\gamma$) is the decay rate of cavity mode $c$ (dopant mode $a$) and $\gamma_{m}$ is the mechanical damping rate. $c_{\rm{in}}$, $a_{\rm{in}}$, and $b_{\rm{in}}$ are the input quantum noises with zero mean values. To solve the nonlinear QLEs in Eq.~(\ref{e04}), we rewrite the operators as the sum of the mean values and the small quantum fluctuation terms, i.e. $c=\xi+\delta c$, $a=\alpha+\delta a$, and $b=\beta+\delta b$. Subsequently, the classical mean values and quantum fluctuations can be treated separately. The equations about steady mean values are given by   
	\begin{eqnarray}\label{e05}
	-(i\Delta_{c}^{\prime}+\kappa)\xi-i[J_{0}+\mu_{0}(\beta^{*}+\beta)]\alpha+\varepsilon&=&0,\cr\cr
	-(i\Delta_{a}+\gamma)\alpha-i[J_{0}+\mu_{0}(\beta^{*}+\beta)]\xi&=&0,\cr\cr
	-(i\omega_{m}+\gamma_{m})\beta-ig_{0}\vert{\xi}\vert^{2}-i\mu_{0}(\alpha^{*}\xi+\xi^{*}\alpha)&=&0,
	\end{eqnarray}
	where $\Delta_{c}^{\prime}=\Delta_{c}+g_{0}(\beta^{*}+\beta)$ is the effective detuning including the frequency shift caused by the mechanical interaction. The solution on $\alpha$ is  
	\begin{equation}\label{e06}
	\alpha=-\frac{i\xi[J_{0}+\mu_{0}(\beta^{*}+\beta)]}{(i\Delta_{a}+\gamma)}.
	\end{equation}
	
	While the linearized fluctuations equations around the steady state can be obtained
	\begin{eqnarray}\label{e07}
	\delta\dot{c}&=&-(i\Delta_{c}^{\prime}+\kappa)\delta c-iJ\delta a-iG(\delta b^{\dag}+\delta b)+\sqrt{2\kappa}c_{\rm{in}},\cr\cr
	\delta\dot{a}&=&-(i\Delta_{a}+\gamma)\delta a-iJ\delta c-i\mu(\delta b^{\dag}+\delta b)+\sqrt{2\gamma}a_{\rm{in}},\cr\cr
	\delta\dot{b}&=&-(i\omega_{m}+\gamma_{m})\delta b-i(G^{*}\delta c+G\delta c^{\dag})-i(\mu\delta a^{\dag}+\mu^{*}\delta a)\cr\cr
	&&+\sqrt{2\gamma_{m}}b_{\rm{in}},
	\end{eqnarray}
	and the linearized effective interactions of the system are clearly depicted in Fig.~\ref{fig1}(c). Hereafter, for simplicity, we define $J=J_{0}+\mu_{0}(\beta^{*}+\beta)$, $g=g_{0}\xi,~\mu=\mu_{0}\xi$, $G=g-iJ\mu/(i\Delta_{a}+\gamma)$, and specify the operators $\delta c\rightarrow c$, $\delta a\rightarrow a$, $\delta b\rightarrow b$. 
	
	For convenience, the linearized QLEs in Eq.~(\ref{e07}) can be written in the matrix form 
	\begin{equation}\label{e08}
	\dot{F}=MF+\Gamma F_{\rm{in}},
	\end{equation}
	where the vectors $F=[c,a,b,c^{\dag},a^{\dag},b^{\dag}]^{T}$ and  $F_{\rm{in}}=[c_{\rm{in}},a_{\rm{in}},b_{\rm{in}},c_{\rm{in}}^{\dag},a_{\rm{in}}^{\dag},b_{\rm{in}}^{\dag}]^{T}$ ($T$ represents the transpose operator), the matrix $\Gamma$= Diag $[\sqrt{2\kappa}, \sqrt{2\gamma}, \sqrt{2\gamma_{m}}, \sqrt{2\kappa}, \sqrt{2\gamma}, \sqrt{2\gamma_{m}},]$, and the coefficient matrix
	\begin{widetext}
		\begin{equation}\label{e09}
		M=\left( 
		\begin{array}{cccccc} 
		-(i\Delta_{c}^{\prime}+\kappa) & -iJ & -iG & 0 & 0 & -iG\\ 
		-iJ & -(i\Delta_{a}+\gamma) & -i\mu & 0 & 0 & -i\mu\\
		-iG^{*} & -i\mu^{*} & -(i\omega_{m}+\gamma_{m}) & -iG & -i\mu & 0\\
		0 & 0 & iG^{*} & -(\kappa-i\Delta_{c}^{\prime}) & iJ & iG^{*}\\
		0 & 0 & i\mu^{*} & iJ & -(\gamma-i\Delta_{a}) & i\mu^{*}\\ 
		iG^{*} & i\mu^{*} & 0 & iG & i\mu & -(\gamma_{m}-i\omega_{m})\\
		\end{array}
		\right). 
		\end{equation}
	\end{widetext}
	The system is stable only if the real parts of all the eigenvalues of matrix $M$ are negative. The stability condition can be derived by applying the Routh-Hurwitz criterion~\cite{PRA74}, however, whose concrete form is too cumbersome to give here. In the following, all the parameters in the present work have been chosen to make sure the stability of the system. 
	
	Generally, it is more convenient to reveal the nonreciprocity in the frequency domain. To this end, we perform the Fourier transformations of the operators
	\begin{eqnarray}\label{e10}
	o(\omega)&=&\frac{1}{2\pi}\int_{-\infty}^{+\infty} o(t)e^{i\omega t}\, dt,\cr\cr
	o^{\dag}(\omega)&=&\frac{1}{2\pi}\int_{-\infty}^{+\infty} o^{\dag}(t)e^{i\omega t}\, dt,
	\end{eqnarray}
	and then Eq.~({\ref{e08}}) in the frequency domain becomes
	\begin{equation}\label{e11}
	F(\omega)=-(M+i\omega I)^{-1}\Gamma F_{\rm{in}}(\omega),
	\end{equation}
	where $I$ represents the identity matrix. By substituting Eq.~({\ref{e11}}) into the standard input-output relation $o_{\rm{in}}+o_{\rm{out}}=\sqrt{\kappa_{o}}o$, where $o_{\rm{in}}$, $o_{\rm{out}}$, and $\kappa_{o}$ denote input operators, output operators, and their corresponding damping rates, respectively. Then the output field vector in the frequency domain is given by
	\begin{equation}\label{e12}
	F_{\rm{out}}(\omega)=U(\omega)F_{\rm{in}}(\omega),
	\end{equation}
	where the output field vector $F_{\rm{out}}(\omega)=[c_{\rm{out}}, a_{\rm{out}}, b_{\rm{out}},c_{\rm{out}}^{\dag}, a_{\rm{out}}^{\dag}, b_{\rm{out}}^{\dag}]^{T}$ and the coefficient matrix $U(\omega)$ is given by
	\begin{equation}\label{e13}
	U(\omega)=-\Gamma(M+i\omega I)^{-1}\Gamma-I.
	\end{equation}
	
	The spectrum of the output field is defined by
	\begin{equation}\label{e14}
	s_{o}^{\rm{out}}(\omega)=\int_{-\infty}^{+\infty}d\omega^{\prime}\langle o_{\rm{out}}^{\dag}(\omega^{\prime})o_{\rm{out}}(\omega)\rangle.
	\end{equation}
	Accordingly, the vector of the output field $S_{\rm{out}}(\omega)$  is written in the form of  $S_{\rm{out}}(\omega)=[s_{c}^{\rm{out}}(\omega), s_{a}^{\rm{out}}(\omega), s_{b}^{\rm{out}}(\omega)]^{T}$. Substituting Eq.~({\ref{e12}}) into Eq.~({\ref{e14}}) and making use of correlation functions $\langle o_{\rm{in}}^{\dag}(\omega^{\prime}) o_{\rm{in}}(\omega)\rangle=s_{o}^{\rm{in}}(\omega) \delta(\omega+\omega^{\prime})$ and $\langle o_{\rm{in}}(\omega^{\prime}) o_{\rm{in}}^{\dag}(\omega)\rangle=[s_{o}^{\rm{in}}(\omega)+1] \delta(\omega+\omega^{\prime})$, we can obtain
	\begin{equation}\label{e15}
	S_{\rm{out}}(\omega)=T(\omega)S_{\rm{in}}(\omega)+S_{\rm{v}}(\omega),
	\end{equation}
	where the vector of the input field $S_{\rm{in}}(\omega)=[s_{c}^{\rm{in}}(\omega), s_{a}^{\rm{in}}(\omega), s_{b}^{\rm{in}}(\omega)]^{T}$ and the transmission matrix
	\begin{equation}\label{e16}
	T(\omega)=\left( 
	\begin{array}{ccc} 
	T_{cc}(\omega) & T_{ca}(\omega) & T_{cb}(\omega)\\
	T_{ac}(\omega) & T_{aa}(\omega) & T_{ab}(\omega)\\
	T_{bc}(\omega) & T_{ba}(\omega) & T_{bb}(\omega)\\
	\end{array}
	\right). 
	\end{equation}
	The element of $T(\omega)$ denotes as $T_{ij}(\omega)$ $(i, j=c, a, b)$, which indicates the scattering probability from the input mode $j$ to the output mode $i$ and their respective forms are
	\begin{eqnarray}\label{e17}
	T_{cc}(\omega)&=&\vert{U_{11}(\omega)}\vert^{2}+\vert{U_{14}(\omega)}\vert^{2},\cr\cr
	T_{ca}(\omega)&=&\vert{U_{12}(\omega)}\vert^{2}+\vert{U_{15}(\omega)}\vert^{2},\cr\cr
	T_{cb}(\omega)&=&\vert{U_{13}(\omega)}\vert^{2}+\vert{U_{16}(\omega)}\vert^{2},\cr\cr
	T_{ac}(\omega)&=&\vert{U_{21}(\omega)}\vert^{2}+\vert{U_{24}(\omega)}\vert^{2},\cr\cr
	T_{aa}(\omega)&=&\vert{U_{22}(\omega)}\vert^{2}+\vert{U_{25}(\omega)}\vert^{2},\cr\cr
	T_{ab}(\omega)&=&\vert{U_{23}(\omega)}\vert^{2}+\vert{U_{26}(\omega)}\vert^{2},\cr\cr
	T_{bc}(\omega)&=&\vert{U_{31}(\omega)}\vert^{2}+\vert{U_{34}(\omega)}\vert^{2},\cr\cr
	T_{ba}(\omega)&=&\vert{U_{32}(\omega)}\vert^{2}+\vert{U_{35}(\omega)}\vert^{2},\cr\cr
	T_{bb}(\omega)&=&\vert{U_{33}(\omega)}\vert^{2}+\vert{U_{36}(\omega)}\vert^{2},
	\end{eqnarray}   
	where $U_{vw}(\omega)$ ($v,w=1,...,6$) is the element in the $v$-th row and the $w$-th column of the matrix $U(\omega)$ in Eq.~({\ref{e13}}). And $S_{\rm{v}}(\omega)=[s_{c}^{\rm{v}}(\omega), s_{a}^{\rm{v}}(\omega), s_{b}^{\rm{v}}(\omega)]^{T}$ in Eq.~(\ref{e15}) is the output spectrum caused by the input vacuum field and its elements are
	\begin{eqnarray}\label{e18}
	s_{c}^{\rm{v}}(\omega)&=&\vert{U_{14}(\omega)}\vert^{2}+\vert{U_{15}(\omega)}\vert^{2}+\vert{U_{16}(\omega)}\vert^{2},\cr\cr
	s_{a}^{\rm{v}}(\omega)&=&\vert{U_{24}(\omega)}\vert^{2}+\vert{U_{25}(\omega)}\vert^{2}+\vert{U_{26}(\omega)}\vert^{2},\cr\cr
	s_{b}^{\rm{v}}(\omega)&=&\vert{U_{34}(\omega)}\vert^{2}+\vert{U_{35}(\omega)}\vert^{2}+\vert{U_{36}(\omega)}\vert^{2}.
	\end{eqnarray} 
	
	\section{RESULTS AND DISCUSSION}\label{sec2}
	\subsection{Nonreciprocal response}
	\begin{figure}
		\includegraphics[width=2.3in]{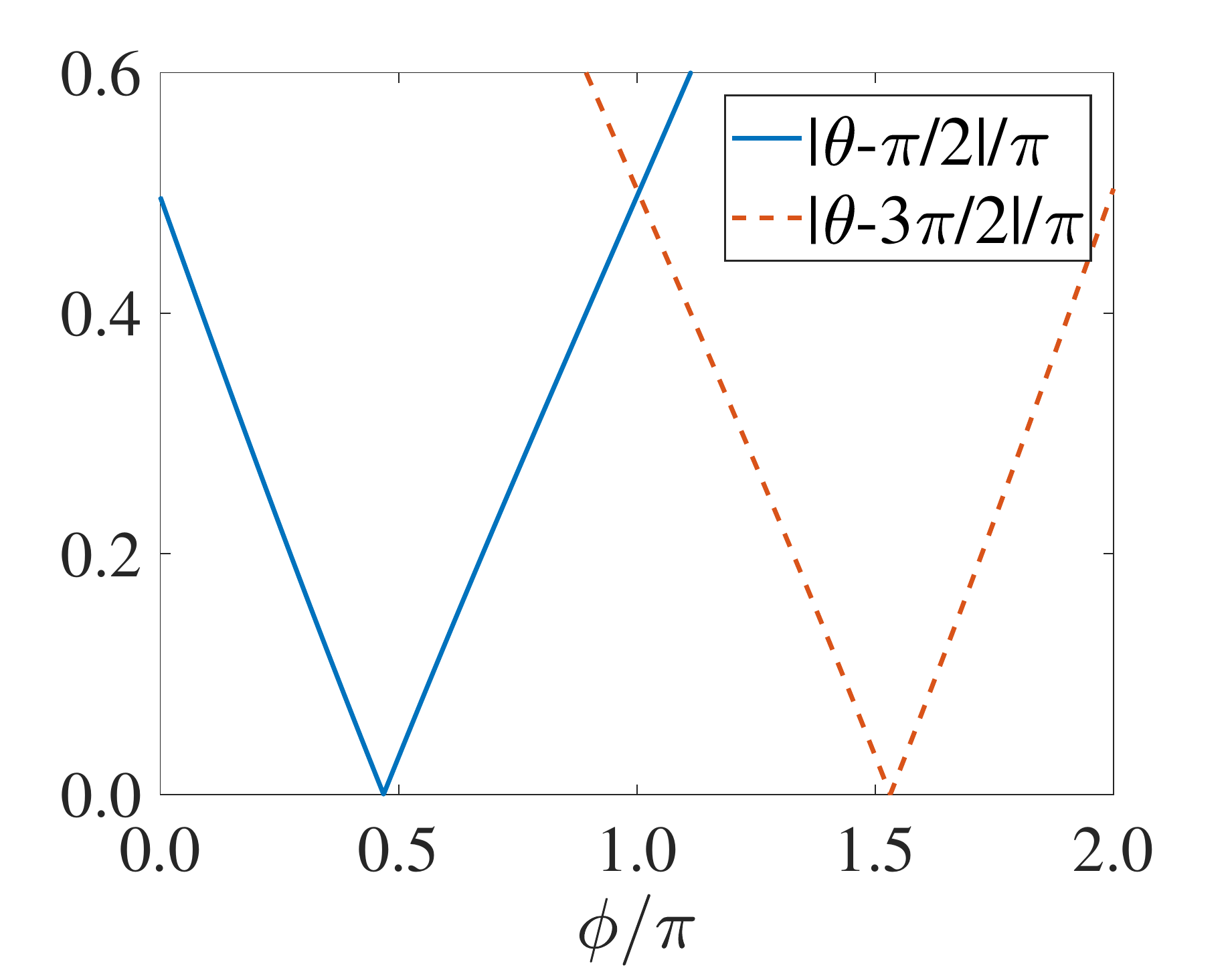}
		\centering
		\caption{(Color online) The relation between the relative phase $\theta$ and the phase $\phi$. The blue solid line represents $\vert\theta-\pi/2\vert/\pi$ while the orange dashed line represents $\vert\theta-3\pi/2\vert/\pi$. The other parameters are $\Delta_{c}^{\prime}=\Delta_{a}=\omega_{m}=10\kappa$ and $g= J=\vert\mu\vert=\gamma=\gamma_{m}=\kappa $.}
		\label{fig2}
	\end{figure}
	
	In this section, we turn to discuss the nonreciprocal transmission between the optical mode and the dopant mode. Based on above definition, it is easy to find that the coupling strength $J$ between the cavity mode and the dopant mode is real while both the effective optomechanical coupling $G$ and interaction strength $\mu$ are complex. Due to the existence of the Tavis-Cummings coupling ($\mu_0\neq0$), as shown in Fig.~\ref{fig1}(c), we cannot simply determine the respective phase at the paths of $B$ and $C$. In the following, the relation of the two phases is derived analytically. If we rewrite $\mu=\mu_{0}\xi=\vert\mu\vert e^{i\phi}$,
	\begin{equation}\label{e19}
	G=g-\frac{iJ\vert\mu\vert(\cos\phi+i\sin\phi)}{i\Delta_{a}+\gamma}=\vert G\vert e^{i\phi^{\prime}},
	\end{equation}  
	where $\phi$ and $\phi^{\prime}$ are the arguments of $\mu$ and $G$, respectively. We specify the relative phase between the paths of $B$ and $C$ is $\theta=\phi-\phi^{\prime}$. Generally, when the relative phase between the two paths is $\pi/2$ or $3\pi/2$, the optimal nonreciprocal response occurs~\cite{PRA39,PRA36,PRA59}. In Fig.~\ref{fig2}, we plot the relative phase $\theta$ with respect to the phase $\phi$, where the parameters are given as $g=J=\vert\mu\vert=\gamma=\gamma_{m}=\kappa$ and $ \Delta_{a}=\Delta_{c}^{\prime}=\omega_{m}=10\kappa$. One can note that the optimal value of the relative phase is $\theta-\pi/2=0$ or $\theta-3\pi/2=0$. Obviously, it does not appear at $\phi=\pi/2$ and $\phi=3\pi/2$, but at $\phi=0.47\pi$ and $\phi=1.53\pi$. Meanwhile, according to the previous calculation and analysis of the optomechanical coupling strength $G$, we can find that the relative phase $\theta=\phi-\phi^{\prime}$ of the two different paths is related to the detuning of the dopant mode $\Delta_{a}$, the interaction strength between the optical mode and the dopant mode $J$, and the effective optomechanical coupling $g$. Therefore, we can modulate the relative phase by adjusting the corresponding system parameters $\Delta_{a}$, $J$ or $g$ for practical system~\cite{NatureC,PRL114093602,PRL105220501}. Next, we discuss the nonreciprocal response when $\phi=0.47\pi$ and $1.53\pi$. 
	\begin{figure}
		\includegraphics[width=4in]{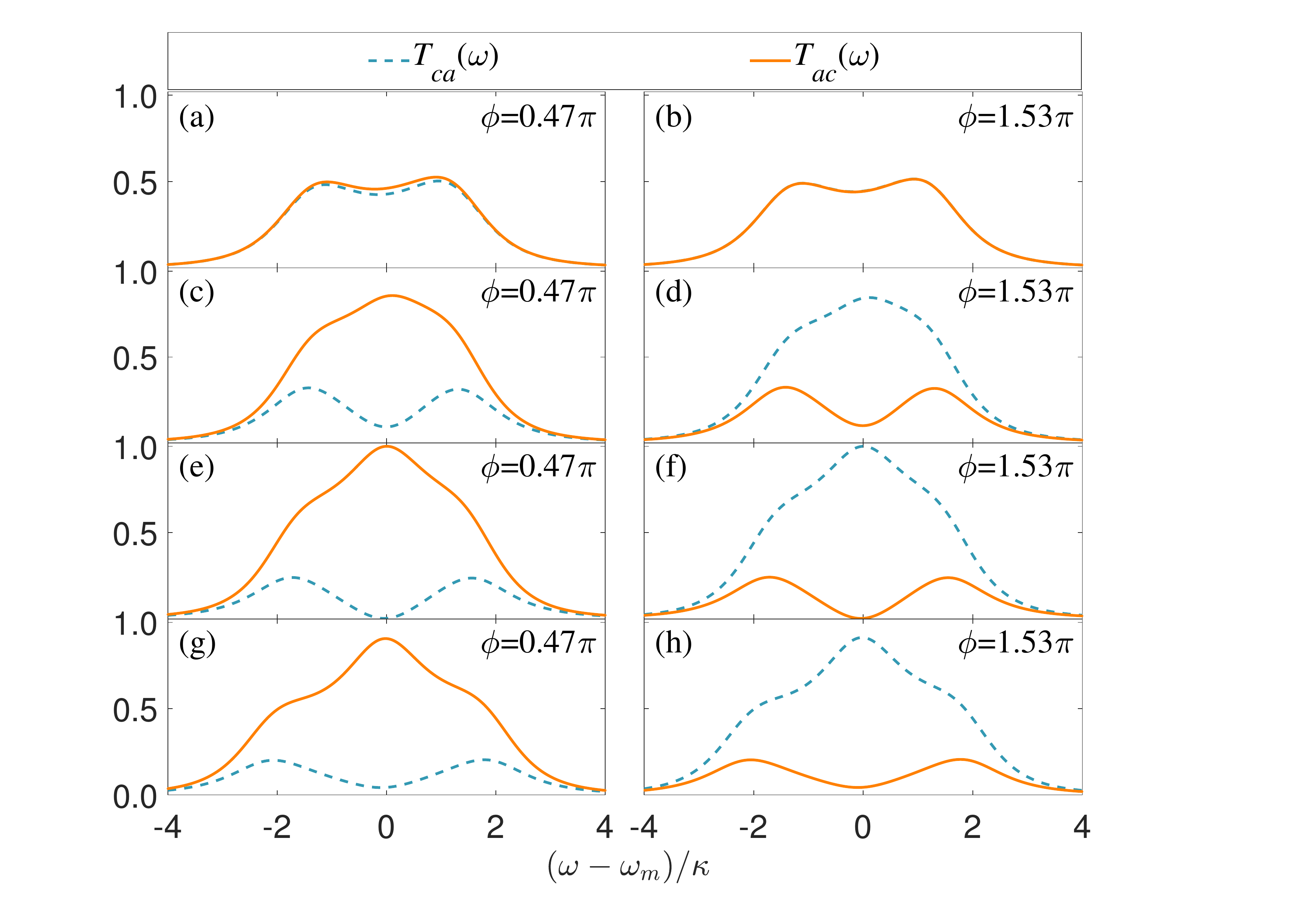}
		\centering
		\caption{(Color online) Scattering probabilities $T_{ca}(\omega)$ (blue dashed line) and $T_{ac}(\omega)$ (orange solid line) as functions of the frequency of the incoming signal $\omega$ for different effective optomechanical coupling rates $g$. These panels reveal the situations of two different phases: (a), (c), (e), and (g) for $\phi=0.47\pi$ while (b), (d), (f), and (h) for $\phi=1.53\pi$. From the top panels to bottom panels, the parameter $g$ is chosen as $0.01\kappa, 0.5\kappa, 1.0\kappa$, and $1.5\kappa$, respectively. The other parameters are the same as in Fig.~\ref{fig2}.}
		\label{fig3}
	\end{figure}
	
	We now focus on how the nonreciprocal response is tuned by the optomechanical coupling strength $g$.  Figure~\ref{fig3} shows the scattering probabilities $T_{ca}(\omega)$ and $T_{ac}(\omega)$ as functions of the frequency of the incoming signal $\omega$ for different coupling rates $g$. It is obvious to find that the scattering probabilities of the mutually inverse processes $T_{ca}$ and $T_{ac}$ reveal opposite nonreciprocal effects at two different phases. When $\phi=0.47\pi$, the scattering probability from mode $a$ to mode $c$ is almost the same as the one from mode $c$ to mode $a$ [i.e., $T_{ca}(\omega)\approx T_{ac}(\omega)$] when the optomechanical coupling strength $g$ is weak (i.e., $g=0.01\kappa$), as shown in Fig.~\ref{fig3}(a). The scattering probability $T_{ac}$ gradually increases with the enhancement of the optomechanical coupling strength $g$ while the opposite direction $T_{ca}$ decreases [i.e., $T_{ac}(\omega)>T_{\rm{ca}}(\omega)$]. Finally, the nonreciprocal response becomes more obvious and the system reaches the optimal nonreciprocal response when $g=\kappa$ at $\omega=\omega_{m}$ [i.e., $T_{ac}(\omega)\approx1, T_{\rm{ca}}(\omega)\approx0$], as shown in Fig.~\ref{fig3}(e). However, if we keep increasing the optomechanical coupling strength $g$, the nonreciprocal response gradually becomes weak. While for $\phi=1.53\pi$, the system exhibits the opposite phenomena.
	
	\begin{figure}
		\includegraphics[width=4in]{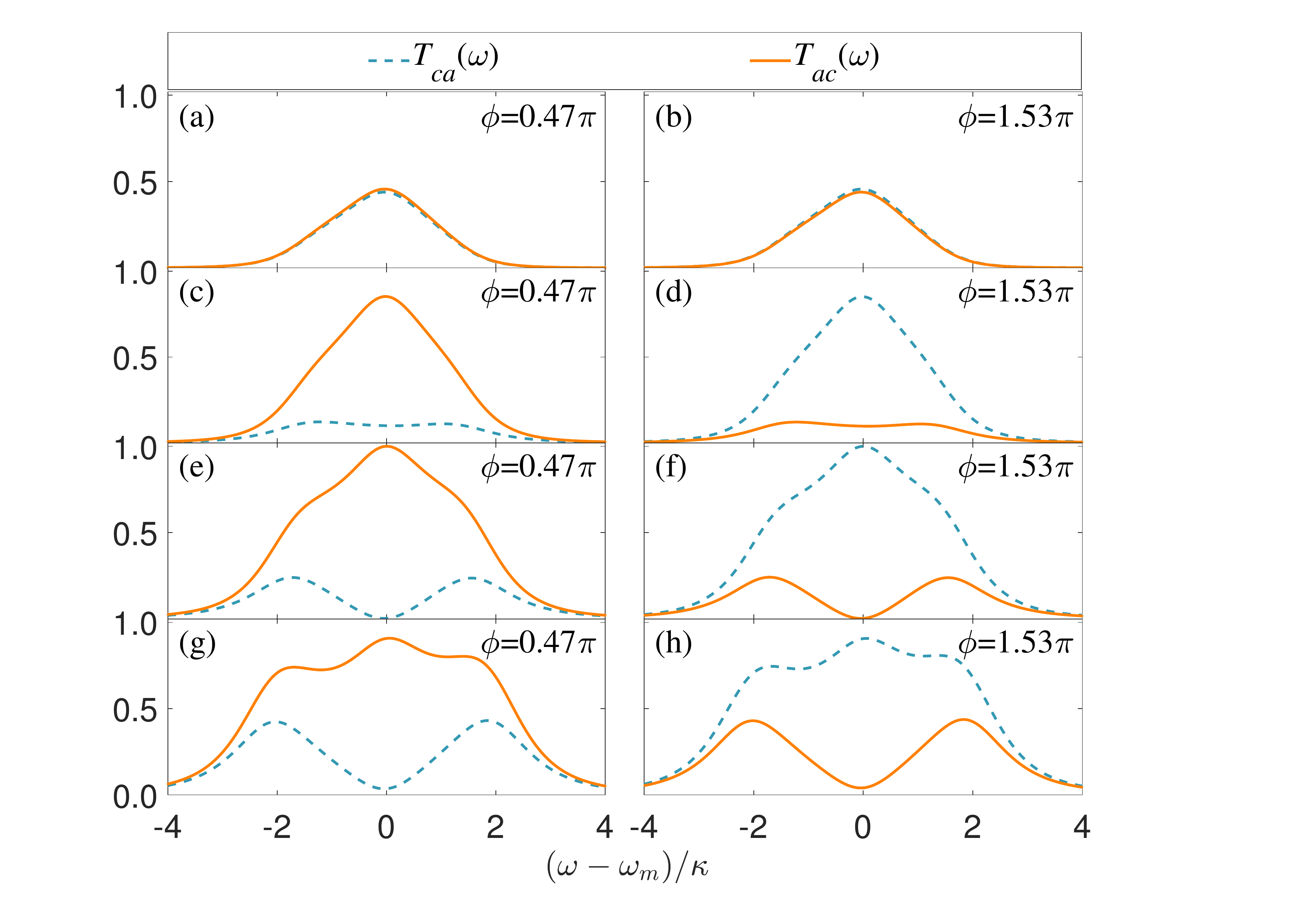}
		\centering
		\caption{(Color online) Scattering probabilities $T_{ca}(\omega)$ (blue dashed line) and $T_{ac}(\omega)$ (orange solid line) as functions of the frequency of the incoming signal $\omega$ for different coupling coefficient $J$. These panels reveal the situations of two different phases: (a), (c), (e), and (g) for $\phi=0.47\pi$ while (b), (d), (f), and (h) for $\phi=1.53\pi$. From the top panels to bottom panels, the parameter $J$ is chosen as $0.01\kappa, 0.5\kappa, 1.0\kappa$, and $1.5\kappa$, respectively. In all  subfigures, $g=\kappa$ and the other parameters are the same as in Fig.~\ref{fig2}.}
		\label{fig4}
	\end{figure}  
	
	In Fig.~\ref{fig4}, the scattering probabilities $T_{ca}(\omega)$ and $T_{ac}(\omega)$ are shown as functions of the frequency of the incoming signal $\omega$ for different interaction strength $J$. It is shown that the scattering probabilities of the two mutually inverse processes $T_{ca}$ and $T_{ac}$ still reveal opposite nonreciprocal effects at $\phi=0.47\pi$ and $\phi=1.53\pi$. When $J\ll\kappa$ (e.g., $J=0.01\kappa$), as depicted in Figs.~\ref{fig4}(a) and~\ref{fig4}(b), $T_{ca}(\omega)\approx T_{\rm{ac}}(\omega)$. In this parameter regime, the system almost does not exhibit the nonreciprocal response. With the increase of the coupling strength $J$, the nonreciprocal response becomes distinct and the system reaches the optimal nonreciprocal effect when $J=\kappa$ at $\omega=\omega_{m}$. As shown in Figs.~\ref{fig4}(g) and~\ref{fig4}(h),  if further increasing coupling strength $J$, the nonreciprocal phenomenon becomes weak,  which is similar with the case in Fig.~\ref{fig3}.
	
	\begin{figure}
		\includegraphics[width=4in]{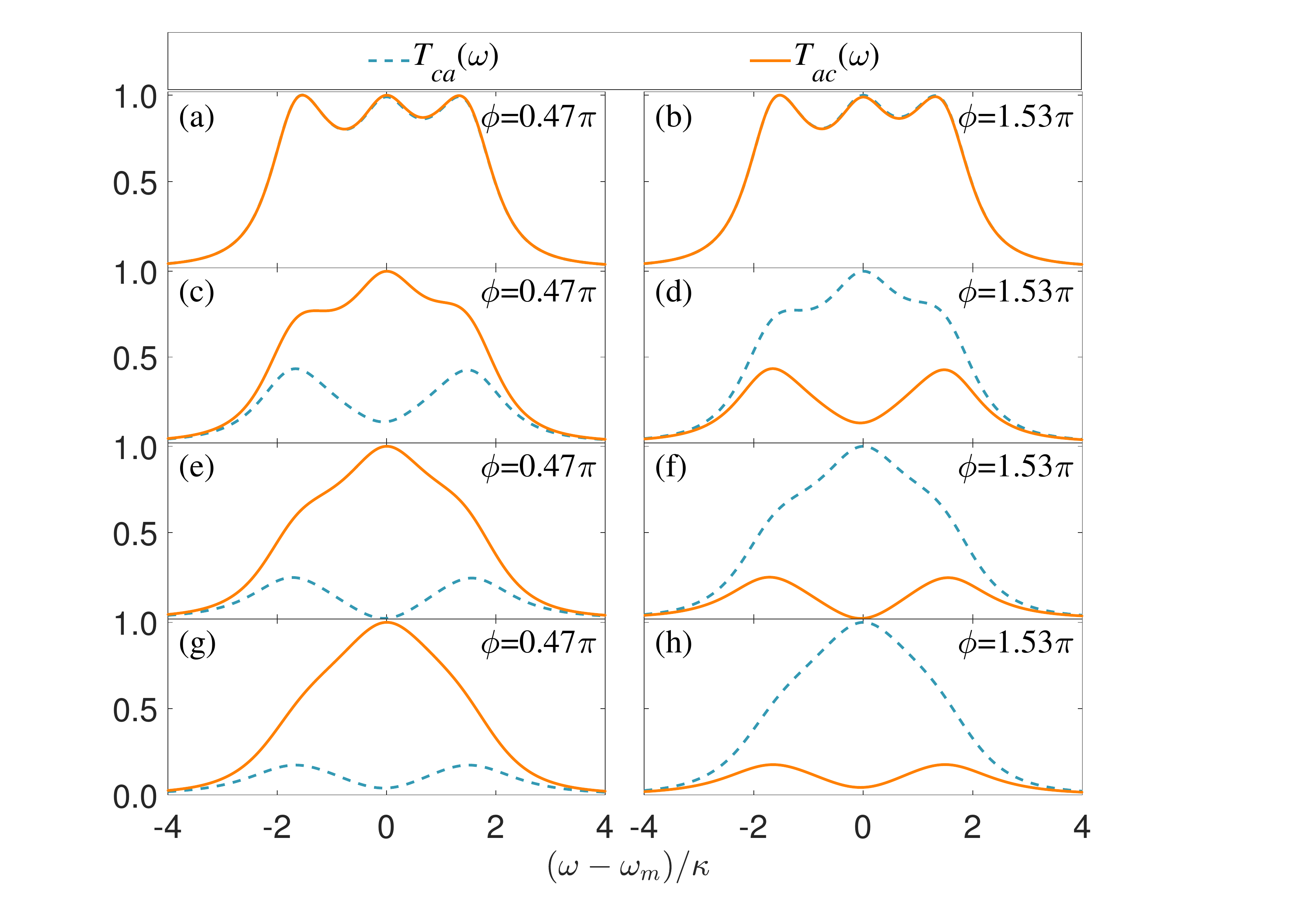}
		\centering
		\caption{(Color online) Scattering probabilities $T_{ca}(\omega)$ (blue dashed line) and $T_{ac}(\omega)$ (orange solid line) as functions of the frequency of the incoming signal $\omega$ for different mechanical damping rates $\gamma_{m}$. These panels reveal the situations of two different phases: (a), (c), (e), and (g) for $\phi=0.47\pi$ while (b), (d), (f), and (h) for $\phi=1.53\pi$. From the top panels to bottom panels, the parameter $\gamma_{m}$ is chosen as $0.01\kappa, 0.5\kappa, 1.0\kappa$, and $1.5\kappa$, respectively. In all subfigures, $g=\kappa$ and the other parameters are the same as in Fig.~\ref{fig2}.}
		\label{fig5}
	\end{figure}
	
	In Fig.~\ref{fig5}, we plot the scattering probabilities $T_{ca}(\omega)$ and $T_{ac}(\omega)$ for different mechanical damping rates $\gamma_{m}$ as functions of the frequency of the incoming signal $\omega$. In the case of $\phi=0.47\pi$, when the mechanical damping rate $\gamma_{m}$ is smaller than the decay rate of the cavity $\kappa$, as shown in Fig.~\ref{fig5}(a), $T_{ca}\approx T_{\rm{ac}}$, which indicates that there is no obvious nonreciprocity. With the increase of the mechanical damping rate $\gamma_{m}$, the nonreciprocal response becomes prominent and the system shows the optimal nonreciprocal effect at $\gamma_{m}=\kappa$, as shown in Fig.~\ref{fig5}(e). However, when $\phi=1.53\pi$, contrary to the case of $\phi=0.47\pi$, $T_{ca}>T_{\rm{ac}}$ occurs once increasing the mechanical damping rate $\gamma_{m}$ and the optimal effect corresponds to $\gamma_{m}=\kappa$ at $\omega=\omega_{m}$. The nonreciprocal response also becomes weak with increasing the mechanical damping rate $\gamma_{m}$ continuously at two different phases. 
	
	\subsection{Phonon-Photon transducer and optomechanical circulator}
	\begin{figure}
		\includegraphics[width=3.3in]{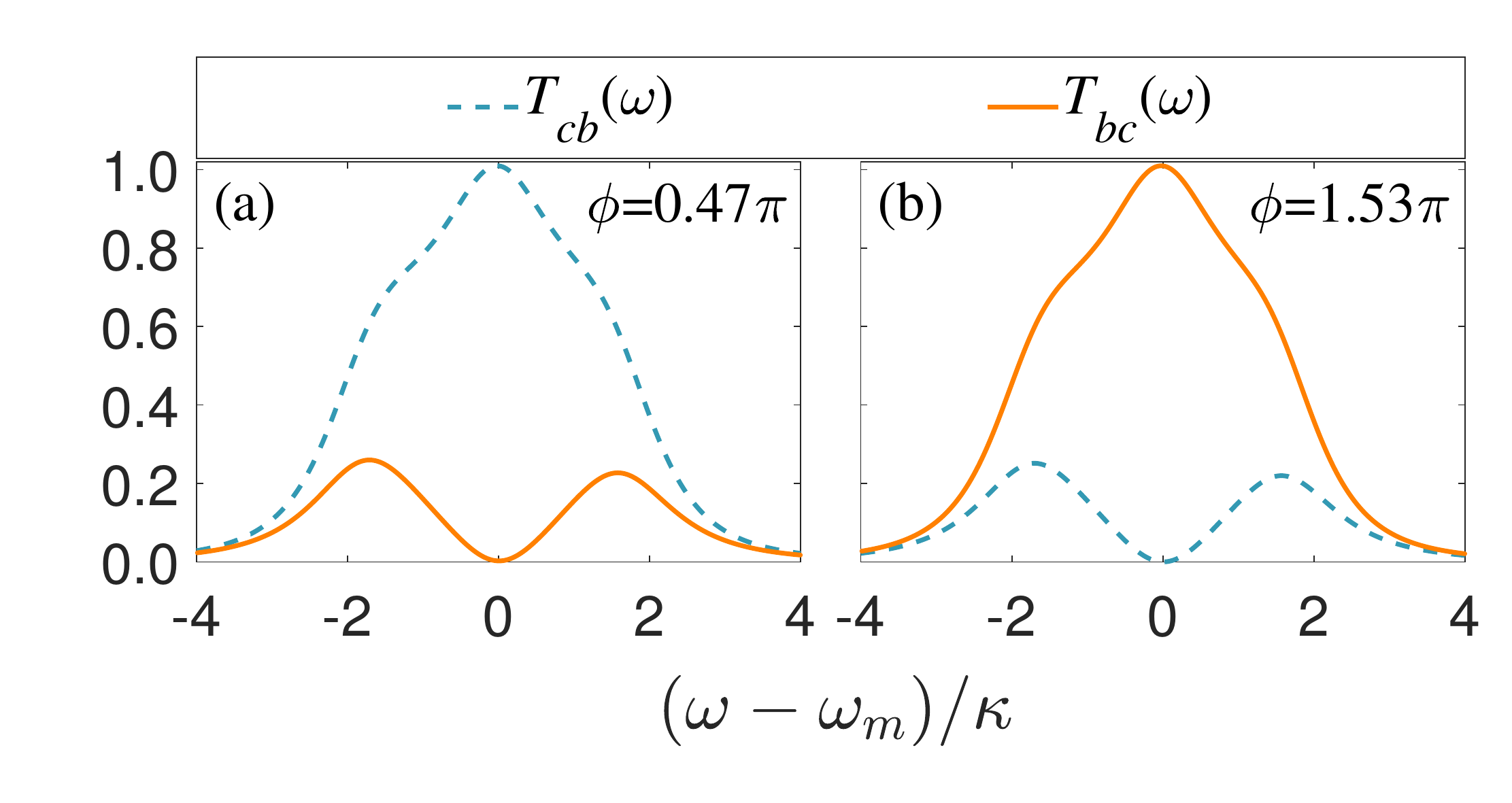}
		\centering
		\caption{(Color online) Scattering probabilities $T_{cb}(\omega)$ (blue dashed line) and $T_{bc}(\omega)$ (orange solid line) as functions of the frequency of the incoming signal $\omega$ for different phases $\phi$. The two panels reveal the situations of two different phases: (a) for $\phi=0.47\pi$ while (b) for $\phi=1.53\pi$. The other parameters are the same as in Fig.~\ref{fig2}.}
		\label{fig6}
	\end{figure}
	\begin{figure}
		\includegraphics[width=4in]{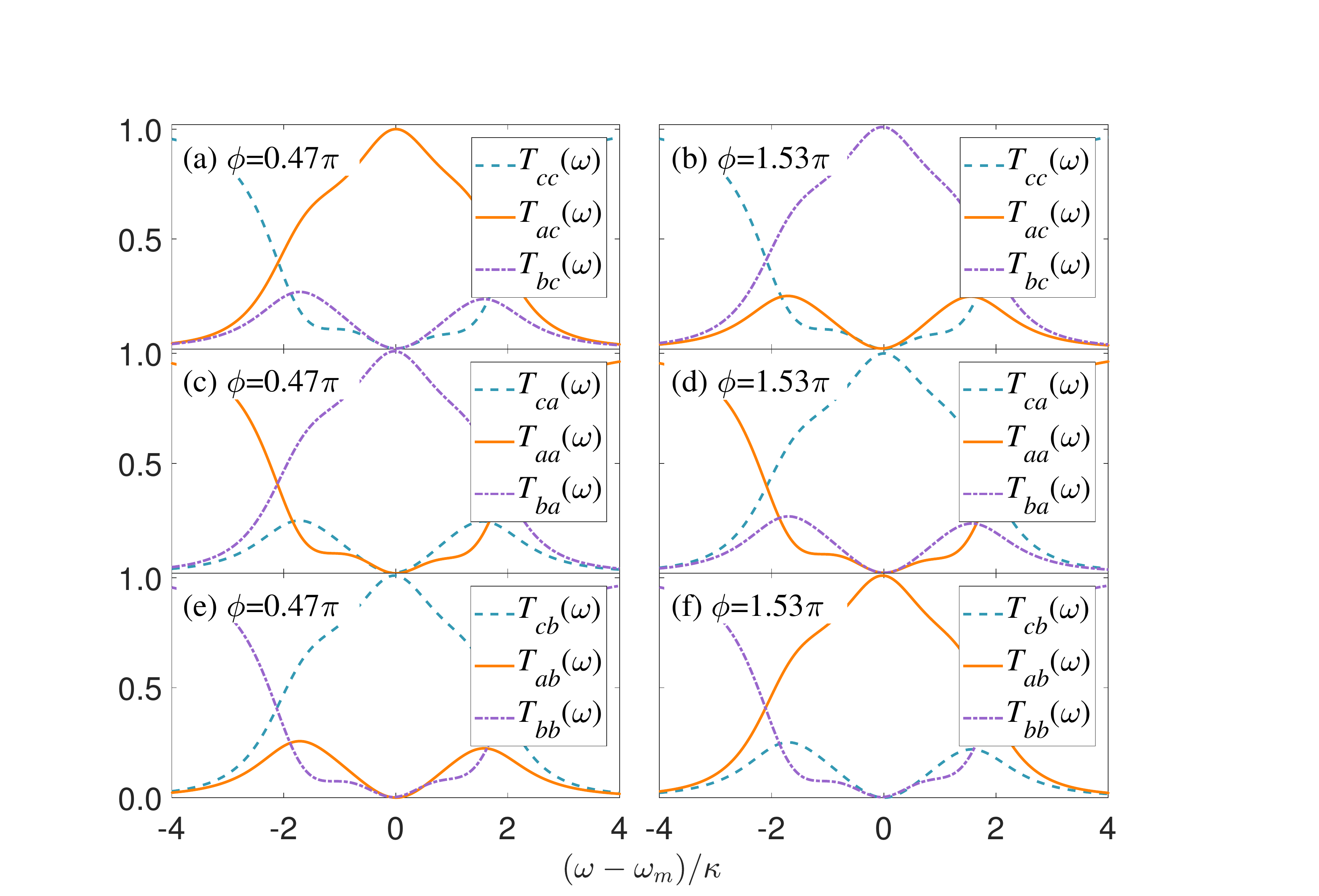}
		\centering
		\caption{(Color online) Scattering probabilities $T_{tc}(\omega)$,   $T_{ta}(\omega)$, and $T_{tb}(\omega)$~($t=c,~a$,~and~$b$) as functions of the frequency of the incoming signal $\omega$ for different phases $\phi$. These panels reveal the situations of two different phases: (a), (c), and (e) for $\phi=0.47\pi$ while (b), (d), and (f) for $\phi=1.53\pi$. The other parameters are the same as in Fig.~\ref{fig2}.}
		\label{fig7}
	\end{figure}
	
	In this section, we show how the proposed Tavis-Cummings coupling optomechanical system can be applied as a photon-phonon transducer and optomechanical circulator. Based on the scattering probabilities $T_{cb}(\omega)$ and $T_{bc}(\omega)$ in Eq.~(\ref{e17}) and the above parameters for producing the optimal nonreciprocal response, as shown in Fig.~\ref{fig6}, we obtain the phonon-photon conversion rates $T_{cb}(\omega)$ and $T_{bc}(\omega)$. It is obvious that the proposed system can realize the nonreciprocal conversion between the optical and mechanical modes, which indicates that the present Tavis-Cummings coupling optomechanical system can be viewed as a phonon-photon transducer and its converted processes can be reversed by changing the phase of the interaction strength.
	
	Figure~\ref{fig7} depicts the scattering probabilities $T_{tc}(\omega)$, $T_{ta}(\omega)$, and $T_{tb}(\omega)$ $(t=c, a, b)$ as functions of the frequency of the incoming signal $\omega$ for different phases based on the optimal nonreciprocal conditions. When $\phi=0.47\pi$, one can notice that $T_{ac}(\omega)\approx T_{\rm{ba}}(\omega)\approx T_{\rm{cb}}(\omega)\approx 1$ but the other scattering probabilities are approximately equal to zero at $\omega=\omega_{m}$, as shown in Figs.~\ref{fig7}(a),~\ref{fig7}(c), and~\ref{fig7}(e). However, as shown in Figs.~\ref{fig7}(b),~\ref{fig7}(d), and~\ref{fig7}(f), once $\phi=1.53\pi$, $T_{bc}(\omega)\approx T_{\rm{ca}}(\omega)\approx T_{\rm{ab}}(\omega)\approx 1$ while the others are approximately equal to zero at $\omega=\omega_{m}$. This indicates that the signal is transferred from one mode to another along a clockwise direction $c\rightarrow a\rightarrow b\rightarrow c$ or counterclockwise direction $c\rightarrow b\rightarrow a\rightarrow c$, which depends on the phase $\phi=0.47\pi$ or $\phi=1.53\pi$. Therefore, the proposed Tavis-Cummings coupling optomechanical system also exhibits the optomechanical circulator behavior and the circulator direction is determined by the phase $\phi$. 
	
	\section{Conclusions}\label{sec3}
	In conclusion, we have studied the nonreciprocal response effect and signal conversion in a Tavis-Cummings coupling optomechanical system. Using the general linearization technique and the Fourier transform, we analytically calculate the transmission matrix of the quantum input signal in the frequency domain. Different from the previous schemes, we find that, due to the introduction of the Tavis-Cummings interaction, both the effective optomechanical coupling strength and the interaction strength between the dopant mode and the mechanical mode are complex and their phases are correlated each other. To obtain the optimal nonreciprocal response, we analytically get the relation of the two phases and derive the relative phase between the two different paths. By selecting the suitable system parameters, especially the relative phase, the optimal nonreciprocal response can be effectively achieved. Furthermore, based on this interesting property, we further discuss the nonreciprocal conversion phenomena among the optical mode, mechanical mode, and dopant mode, which can be used as a phonon-photon transducer and an optomechanical circulator. We hope that the proposed proposal would have potential applications in quantum information processing network.

	\begin{center}
		{\small {\bf ACKNOWLEDGMENTS}}
	\end{center}
	
	This work was supported by the National Natural Science Foundation of China under Grant Nos. 61822114, 61465013, and 11465020.

\end{document}